# Machine Learning Applications in Misuse and Anomaly Detection

Jaydip Sen* & Sidra Mehtab
School of Computing and Analytics, NSHM Knowledge Campus, Kolkata, India.
*Corresponding author email: jaydip.sen@acm.org.

## Abstract

Machine learning and data mining algorithms play important roles in designing intrusion detection systems. Based on their approaches towards the detection of attacks in a network, intrusion detection systems can be broadly categorized into two types. In the misuse detection systems, an attack in a system is detected whenever the sequence of activities in the network match with a known attack signature. In the anomaly detection approach, on the other hand, anomalous states in a system are identified based on a significant difference in the state transitions of the system from its normal states. This chapter presents a comprehensive discussion on some of the existing schemes of intrusion detection based on misuse detection, anomaly detection and hybrid detection approaches. Some future directions of research in the design of algorithms for intrusion detection are also identified.

**Keywords:** misuse detection, signature detection, anomaly detection, hybrid detection, clustering, classification, unsupervised learning, supervised learning, intrusion detection system, network security.

## 1. Introduction

Cyberinfrastructures are vulnerable to various possible attacks due to the flaws in their design and implementation. The major flaws that cause most of the critical vulnerabilities are errors in system programs and faulty design of the software. Malicious attackers can exploit these system vulnerabilities by following a sequence of activities, either from inside or from outside of the infrastructure, and cause significant damage. These events manifest themselves in the form of different distinct characteristics that are defined as patterns of attacks. Misuse or signature detection techniques attempt to proactively detect the presence of such patterns so that any malicious attack on the infrastructure can be effectively defended against. It is possible to defend against all known vulnerabilities in cyberinfrastructures by using supervised learning approaches for misuse and signature detection. The most convenient method of signature detection is measuring the similarity between the patterns recognized in the current network activity and the already known patterns of various types of cyber-attacks. However, execution signatures may vary substantially from one attack category to another, so that specific detection methods are required to classify attack patterns and thus, to improve detection capability.
Anomaly detection systems, however, work in a different way. The objective of these systems s to proactively detect any activity or event in a network or host computer that exhibits aberration from the normal behavior of the network or the host. The normal behavior is described by a predefined set of activities. The working principle of an anomaly detection system is fundamentally different from that of misuse or signature detection system. Misuse or signature detection systems first need to be equipped with a well-defined set of attack signatures populated in its database. An anomaly detection system, on the other hand, defines a detailed and accurate profile of the normal behavior of the networks and hosts. The normal state of the cyberinfrastructure, consisting of networks and hosts, indicates an attack-free state. When an anomalous activity occurs in the cyberinfrastructure, the anomaly detection system notices a state change from the normal state to a state that is no longer normal. On observing this state change, the anomaly detection system raises an alert of a possible attack on the cyberinfrastructure. Unlike the signature or misuse detection systems, the anomaly detection systems are capable of detecting novel attacks as the detection strategy for these systems is based on the state change information, rather than a matching of attack signatures. It is precisely for this reason that anomaly detection schemes are capable of detecting various different types of attacks. Some of these attacks include: (i) segmentation of binary code in a user password, (ii) backdoor service on a malicious process on a well-known port number in a computing host., (iii) stealthy reconnaissance attempts, (iv) novel buffer overflow attacks, (v) direction of hypertext transmission (HTTP protocol on a non-standard port number, (vi) stealthy attacks on protocol stacks, (vii) different variants of denial of service (DoS) and distributed denial of service (DDoS), and so on. Early and accurate detection of these attacks poses significant challenges in the design of a robust and accurate anomaly detection system.

In this chapter, we have briefly reviewed some of the well-known misuse and anomaly-based detection systems that are proposed in the literature. We have also discussed some hybrid approaches in intrusion detections that effectively combine misuse and anomaly detection approaches so as to improve the detection accuracy and reduce false alarms. The rest of the chapter is organized as follows. Section 2 presents a brief discussion on misuse or signature-based detection approach. In Section 3, we discuss how various machine learning approaches can be applied in misuse or signature-based systems. Section 4 provides a brief overview of anomaly detection, while in Section 5 we discuss how machine learning and data mining algorithms can be effectively deployed in anomaly-based detection systems. In Section 6, we briefly discuss the working principles of some of the well-known hybrid detection systems. Section 7 concludes the chapter while highlighting some of the recent trends in machine learning approaches in network security applications.

## 2. Misuse or Signature Detection

Misuse detection, also called signature detection, is an approach in which attack patterns or unauthorized and suspicious behaviors are learned based on past activities and then the knowledge about the learned patterns is used to detect or predict subsequent similar such patterns in a network. The attack or misuse patterns, which are also called signatures, include patterns of log files or data packets which were found to be malicious and identified as threats to the network and the computing hosts. Each log file consists of its own signature that exhibits a unique pattern consisting of binary bits 0 and 1. For intrusion detection systems protecting host computers, i.e., for host-based intrusion detection systems (HIDSs), the attack signature databases may contain various patterns of system calls which represent a different attack on the host. In the case of a network-based intrusion detection system (NIDS), attack signatures reveal specific patterns in data packets. These patterns may include signatures of the data payload, the packet header, unauthorized activities, such as improper file transfer protocol (FTP) initiation, or failed login attempt in

Telnet. A typical data packet includes several fields such as: (i) the source Internet protocol (IP) address, (ii) the destination IP address, (iii) the source port number for transmission control protocol (TCP) or user datagram protocol (UDP), (iv) the destination port number for TCP or UDP, (v) the protocol description such as UDP, TCP or Internet control message protocol (ICMP), and (vi) the data payload. An attack signature can be detected in any specific field, or in any combination of these fields.

Figure1 shows how a typical misuse or signature detection system works. These detection systems execute algorithms that attempt to match learned patterns or signatures from past attacks with the current activities in a network in order to detect any possible attack or malicious activities. If the signature of any current activity in the network matches with the signature of any activity in the attack signature database, the detection system raises an alert. A module in the detection system initiates a further investigation of the attack and starts invoking appropriate security modules to defend against such attacks. If the attack is found to a real attack and not a false alarm by the detection system, the existing database of the attack signatures is updated with the signature of the new attack. For example, if the signature of an attack is: *login name = "Sidra"*, then, whenever there is any attempt to login into any device in the network with the name "Sidra", the signature detection system will raise an alert of an attack.

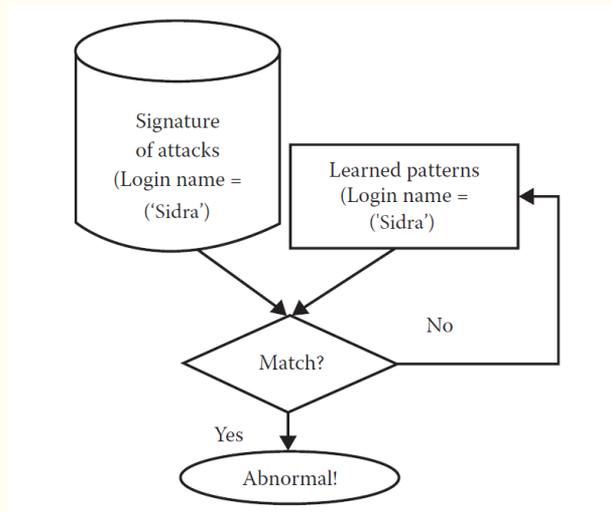

Figure 1: Working of misuse or signature detection: Illustration of "if-else" rules

This approach adopted in a signature-based detection system is primarily meant for detecting already known threats and vulnerabilities in a network. However, these systems suffer from a drawback of producing too many false alarms. A false alarm or a false positive refers to a situation where the system raises an alert of an attack while no attack has really happened on the network. As an example, let us consider the case where a user logs into a remote server. If the user forgets the login password and makes multiple attempts of login, the account of the user is most likely to be locked after a certain number of such failed attempts. As the signature-based detection system cannot differentiate between a failed login attempt by a legitimate user, and a malicious user attempting to login in an unauthorized way into some legitimate user's account, both the activities are considered as attacks.

The efficacy of misuse or signature detection system largely depends on the completeness and sufficiency of the knowledge of attack patterns and signatures captured in the attack signature database of the system. It is a non-trivial task to capture and represent the knowledge of attacks and system vulnerabilities in a cyberinfrastructure or in a network of computing machines, and the job heavily depends on domain experts. Since the knowledge and skills of domain experts may vary significantly from person to person, the design of signature detection systems, quite often, can be incomplete and inaccurate. Moreover, a slight variation, evolution, blending, or a combination of already known attacks can make signature detection an impossible task. This is a typical problem with any similarity-based learning system like a signature-based intrusion detection system.

## 3. Machine Learning in Misuse or Signature Detection

Figure 2 depicts the working mechanism of misuse or signature detection consists of five major steps: (i) data collection, (ii) data pre-processing, (iii) misuse or signature identification using a matching algorithm, (iv) rules regeneration, and (v) denial of service (DoS) or other security response strategy. In most of the cases, the data sources are: network and host audit logs, packets transmitting over the network, and windows registry. Data pre-processing is a critical step that prepares the raw data for learning patterns. These steps involve the reduction of noise by eliminating outliers, normalizing or standardizing of data, and finally selecting and extracting features. After the data pre-processing step is over, an automatic intelligent learning system is deployed to build a learning model and extract rules using prior knowledge of the execution of malicious programs, network traffic data, and vulnerabilities in network infrastructure. The model is now ready for signature and misuse detection. The learned classification model is applied to the incoming network traffic for signature detection. If any part of the network traffic is found to be similar to attack patterns learned by the model, then an alarm is raised and the traffic is further analyzed for identifying whether it is really an attack or a false alarm. Consequently, misuse or signature detection can be simply understood as an "if-then "sequence as shown in Figure 1.

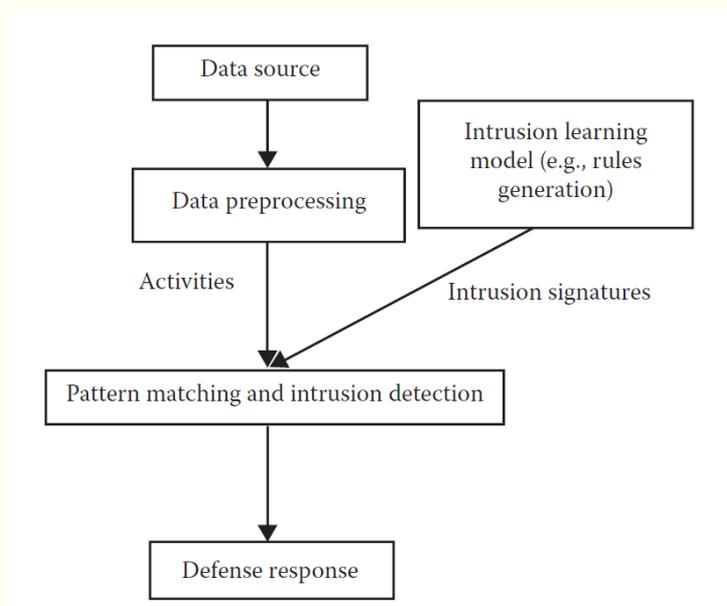

Figure 2: Sequence of execution of misuse or signature detection modules

We present a variety of misuse techniques that are based on machine learning methods. In the following, we discuss some examples of machine-learning methods applied in misuse detection systems.

*Classification using Association Rules*: Agrawal et al. proposed an elegant approach to discover underlying association rules to identify and then establish causal relationships among attributes that may exist in a multidimensional database [1]. Association rules mining identifies the frequent existing patterns in a dataset. This may help, for example, in designing algorithms for a computer antivirus software. A computer antivirus attempts to identify viruses that exhibit some frequently occurring patterns in a transaction dataset. The use of association rules mining and frequently occurring episodes from the computer audit data and exploiting those rules in feature selection had also been described in the literature [2]. Fuzzy association rules were designed for misuse and signature detection on 1998 DARPA intrusion detection dataset [3]. For the purpose of feature selection, 41 features were extracted for each connection record that included 24 attack different attack types. The attack traffic in the network was essentially of four types: (i) denial of service (DoS), (ii) remote to user (R2L), (iii) user to root, and (iv) probes. Including the normal traffic in the network, the association rule mining algorithms extracted the essential features of five types of network data – four categories of attack traffic and one type of normal traffic.

*Artificial Neural Networks:* In a connectionist approach, ANNs carry out the task of pattern recognition and pattern matching tasks using very complex nonlinear transformation functions and the use of multiple hyperplanes separating data of one class from the other. The dynamic nature of the network traffic and the ever-changing characteristics of various attacks on the networks require a very flexible and adaptive misuse detection system that can efficiently and effectively identify a variety of intrusions. Application of ANNs in designing a misuse detection system incorporates the ability to analyze data even if the data may be noisy, distorted, and incomplete. Since ANNs have the ability of learning very accurately from training data, these models can very effectively detect misuse attacks and identify suspicious events in a network. However, this hypothesis is based on the assumption that attackers usually deploy the same approach of an attack on multiple networks, and ANNs can effectively detect similar attacks that had been used by the attackers in the past. Use of ANNs for misuse and signature-based intrusion detection is discussed in [4]. The intrusion detection system (IDS) presented by the author exploits the ability of ANN in classifying nonlinearly separable data into various classes even if the data sources are noisy and limited. The ANN, which is also called a feed-forward multi-layer perceptron (MLP), is equipped with four fully connected layers of nodes with nine nodes in the input layer, two nodes in the output layer, and two hidden layers between the input and the output layers. The two nodes in the output layer are used to indicate the classification results of the network traffic – the normal traffic data being classified with a label of 1, and the attack traffic with a label of 0. Nine important features were extracted by the ANN from the network event data. The network event data are gathered from the data packets transmitted over the network. The nine features of network data traffic that are extracted by the ANN are: (i) protocol identified, (ii) source port number, (iii) destination port, (iv) source internet protocol (IP) address, (v) destination IP address, (vi) internet control message protocol (ICMP) type, (vii) ICMP code, (viii) data payload length, and (ix) data payload content. Each record of network traffic data was first pre-processed, its features were extracted, and then the features having categorical values were transformed into some standardized numeric values. Around 10,000 network traffic records were synthetically generated for the training and the testing of the ANN model, and approximately 3000 among those records were detected to be anomalous.

*Support Vector Machines:* Because of its intrinsic characteristics, support vector machines (SVMs) are capable of minimizing the structural risk of a dataset by reducing its classification error on unseen records, unlike ANNs which focus more on minimization of empirical risk of the dataset. In order to achieve its goal, a model based on the SVM approach determines its number of parameters based on the margin that separates the data points. This margin is determined by the number of support vectors present in the dataset. The support vectors are those data points which lie nearest to the hyperplane but belong to different classes. In contrast to an ANN, the number of parameters in an SVM model dosed not depend on the number of feature dimensions in the dataset. This unique property of an SVM makes it so powerful in many practical machine learning applications. In the context of intrusion detection applications, SVMs present two distinct advantages over their ANN counterparts. SVM models execute much faster and they are more scalable. High speed in execution is crucial for detecting attacks in real-time, while scalability is a mandatory requirement for deployment in a complex cyberinfrastructure. Moreover, SVM models can be made to adapt fast based on changes introduced in the training dataset. This feature of SVM is critical when the patterns in the attack traffic change very rapidly. Mukkamala et al. demonstrated how SVM models can be deployed for the purpose of detection of an attack and misuse patterns in context to computer security breaches [5]. The security breaches considered by the authors were bugs in system software bugs, hardware or software failures, incorrect system administration procedures, or failure of the system authentication. For the purpose of building the SVM model, the authors used a training set of 699 data points that contained some records representing actual attack traffic, some records that represented probable attacks, and remaining records exhibiting normal traffic patterns. Eight features were extracted after the initial cleaning and pre-processing phase of the data. Finally, all feature values for each record were normalized to [0,1]. The test dataset consisted of 250 data points and eight features. In the confusion matrix yielded by the classification model produced a precision value of 85.53% on the training data set, and the corresponding value for the test dataset was 94%. This experiment clearly demonstrated the fact that SVM is, in general, more efficient and accurate in identifying misuse and signature-based attack traffic than its ANN counterpart. It also validated the hypothesis that an SVM can effectively simulate security scenarios using its component to adapt to a given information system. Once adapted to a given system, an SVM model can carry out real-time detection of attack traffic, and minimize false alarms while yielding a very high true detection rate.

*Decision Tree and Classification and Regression Tree:* Decision tree, is a non-parametric machine-learning method of model building that does not impose any preconditions or requirements on the data. A typical decision tree uses a classification algorithm that labels a data point based on the feature values in the data record corresponding to that node in the decision tree. In order to arrive at a classification decision corresponding to a leaf node in a decision tree, one has to trace the path from the root node to the leaf node. The trace of the path from the root node to a given leaf node can then be converted to a classification rule. If designed optimally, decision trees can yield high classification accuracy, while they involve less complexity in implementation, and have the ability to model intuitive knowledge stored in a high dimensional dataset. It is precisely these characteristics that make decision trees a very popular choice in many real-world applications. Among the decision-tree algorithms, CART represents trees in a form of binary recursive partitioning. It classifies objects or predicts outcomes by selecting from a large number of variables. The most important of these variables determine the outcome variable. Kruegel and Toth proposed a signature and misuse detection system following a decision tree-based approach [6]. In the scheme proposed by the authors, the original rules were partitioned into a smaller subset of rules in such a way that the analysis of a single subset is enough for each input element in the signature detection system [6]. The decision tree algorithm was utilized for detecting the feature that most effectively discriminated against the rule sets of different classes. The algorithm is executed in parallel for evaluating each feature on all the rules in a subset. In the decision tree, the root node corresponded to the universal set of rules. In other words, the root node contained all the rules. The children nodes represented the direct subsets of rules that were partitioned from the rule set based on the first feature in the dataset. The splitting of the nodes in the tree continued till a stage was reached here each node was found to contain one rule only. Labeling was done on each node using the feature that was used for splitting the node. Each directed edge emanating from a node and impinging on its child was marked with the value of the feature specified in the child node. Each leaf node contained either one rule or a set of rules that were not distinguishable by the features in the dataset. During splitting, the sequence of features encountered had an impact on the shape and depth of the tree structure. The authors had also proposed an algorithm that generated a decision tree for detecting malicious events using a limited number of comparisons on the set of rules extracted. Chebrolu et al. used KDD cup 1999 intrusion detection data set to build a classification and regression tree (CART) [7]. The dataset included 5092 cases and 41 variables. There were 208,772 possible splits in the CART algorithm. Gini index was used for determining the optimal splitting at the nodes.

*Bayesian Network Classifier:* The major shortcoming of most of the rule-based approaches to classification is that these methods treat each event in isolation and never consider the entire gamut of events together taking into account their contextual and temporal relationships. A rule is derived based on the signature of a packet. The signature of a packet is determined using a set of protocols. Many a time, the signature exhibited by a subset of packets belonging to the activities of a malicious user may match to that of a normal user, rule-based misuse detection systems often suffer from high rates of false alarm. In the case of a false alarm, the intrusion detection system erroneously identifies an activity in a network as malicious while the activity is actually perfectly normal. Bayesian Network (BN)-based models get rid of this problem of rule-based detection systems. Using Bayesian statistics BN represents problems in networks by specifying the causal relationships between subsets of variables. Typically, a BN is presented as a directed graph that does not contain any cycle. Hence, a BN is also referred to as a DAG – directed acyclic graph. Each node in a BN represents a random variable. A random variable is a variable that can assume a set of values; each value has a specified probability of occurrence. Each arc in a BN depicts a causal relationship with the dependence of the child node on the parent node being expressed as a conditional probability value. The head node and the tail node of an arc are referred to as the parent node and the child node respectively. For example, if in a BN, there is an arc $X_1 \rightarrow X_4$, then $X_1$ is the predecessor of node $X_4$, and $X_4$ is the descendent of node $X_1$. In the example BN, the node $X_1$ has no predecessor. However, it has three descendant nodes: $X_2$, $X_3$, and $X_4$. Along with BN, the conditional probability table (CPT) presents the dependencies on the net for each variable/node. For each variable/node, the conditional probability $P$(variable | parent (variable)) is given in CPT for each possible combination of its parents [8-10]. Chebrolu et al. investigated the performance of a feature selection and classification algorithm using BN [7]. Markov Blanket method was used to find the most significant feature set in a training dataset that included five classes of network traffic: normal, probe, DOS, U2R and R2L.

*Naïve Bayes:* The naïve Bayes classifier makes the assumption of class conditional independence. Given a data sample, its features are assumed to be conditionally independent of each other. This is in contrast with a BN that assumes dependencies among the features. Schultz et al. used the naïve Bayes approach to detect new, previously unseen malicious executables accurately and automatically [11].

Most of the machine learning methods for misuse and signature detection are in the initial stages of research and are yet to find any commercial deployment. Moreover, feature selection before traffic classification is a challenging task. Detection quality heavily depends on the experience and knowledge of the security experts dealing with the problem. It also depends on an exhaustive testing and refining process. The use of decision trees for the selection of a significant feature subset has only partially solved this problem. Table 1 summarizes the signature-based detection schemes we discussed in this Section. We have categorized the schemes based on their approach, input data used and level of detection.

**Table 1: Misuse or Signature-Based Detection Schemes**

| Detection Mechanism | Input Data Format | Detection Level | References |
|---|---|---|---|
| Rule-based signature detection | Frequency of system calls, offline | Host | [2] |
| Fuzzy association rules | Frequency of system calls, online | Host | [3] |
| Artificial Neural Networks | TCP/IP packets, offline | Host | [4], [20] |
| Support Vector Machines | TCP/IP packets, offline | Network | [5] |
| Linear genetic programs | TCP/IP packets, offline | Network | [3] |
| Decision tree | TCP/IP packets, online | Network | [6] |
| Classification and regression trees | Frequency of system calls, offline | Host | [7] |
| Statistical method | Executables, offline | Host | [11] |
| Bayesian networks | Frequency of system calls, offline | Host | [7] |

## 4. Anomaly Detection

When a novel attack is launched on a network, misuse detection systems cannot detect the attack as the attack signature is not present in the existing database of attack signatures. However, an anomaly detection system has the ability to detect new and unseen attacks and raise an early alarm before and substantial damage to the network could be done by the attack. Like the misuse detection approach, anomaly detection relies on determining a clear boundary between the normal and the anomalous traffic. The profile of the normal behavior is assumed to be significantly from that of the anomalous behavior. The profile of the normal events and the normal traffic should preferably satisfy a set of criteria in the sense that it must contain a very clearly defined normal behavior. For example, the normal behavior specification must include the IP address or the hostname of a computing machine, or it should include a virtual local area network (VLAN) details to which it belongs, and have the ability to track the normal behavior of the target environment sensitively. In addition, the normal profile should include the following details: (i) occurrence patterns of some specific system calls in the application layer of the communication protocol stack, (ii) association of data payload with different fields of application protocols, (iii) connectivity patterns between secure servers and the Internet, and (iv) the rate and the burst length distributions of all traffic types [12]. In addition, profiles based on a network must be adaptive and self-learning in complex and challenging network traffic to preserve the accuracy and a low false acceptance rate (FAR).

In a large data network, detection of malicious and anomalous traffic is a complex task that poses some significant critical challenges. It is difficult to analyze and monitor a huge volume of traffic that contains network data with a very high-dimensional feature space. Such monitoring and analysis of network traffic data class for highly efficient computational algorithms in data processing and pattern learning. Moreover, the anomalous traffic in a network exhibit a common behavior. In a large volume network traffic data, the malicious and anomalous traffic of the same type tends to occur repeatedly, while the number occurrences of malicious and anomalous data are much smaller than the number of occurrences of normal data. This makes the network traffic data highly imbalanced. It is also difficult, if not impossible, to determine accurately a normal region, or define the boundary between the normal and the anomalous traffic. To complicate the issue further, the concept of anomaly varies among different application domains. In many situations, labeled anomalous data are not available for the training and validation processes. Training and testing data contain the noise of unknown distributions, and the normal and anomalous behavior constantly changes. All these issues make anomaly detection in a network a particularly difficult task.

## 5. Machine Learning in Anomaly Detection

Figure 3 depicts the schematic diagram of a typical anomaly detection system. Anomaly detection systems broadly work in the five steps: (i) data collection, (ii) data pre-processing, (iii) normal behavior learning phase, (iv) identification of misbehaviors using dissimilarity detection techniques, and (v) security responses. In a large-scale network, the data collection phase involves a large volume of data to be collected from the network. In the data pre-processing phase, the volume of data is reduced as this step includes feature selection, feature extraction, and finally dimensionality reduction processes.

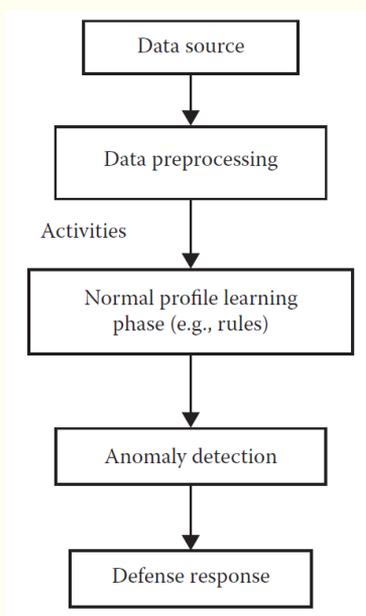

Figure 3: Sequence of execution of modules in an anomaly detection system

Machine-learning algorithms can be very effective in building normal profiles and then designing intrusion detection systems based on anomaly detection approach. In the anomaly detection approach, the network traffic data belonging to a normal class are usually available for training the model. However, in most of the applications, labeled data for anomalous traffic are not available. We have already seen that supervised machine learning algorithms need attack-free training data. In other words, supervised learning needs labeled network data for both types of traffic – normal and attack. However, in most of the real-world situations, such pre-labeled training data for both classes are very difficult to get. In most cases, not only the pre-labeled training data are not available, the traffic data in networks exhibit highly imbalance characteristics. A large majority of normal traffic record is mixed with a tiny minority of attack traffic records. To make the challenge even bigger, with the change in the network environment, patterns of normal traffic also exhibit substantial changes. The significant difference in the characteristics of training and test datasets most often leads to high false positive rates (FPRs) for supervised intrusion detection systems (IDSs). Unsupervised learning methods as adopted by anomaly detection systems can potentially get rid of this problem by building a normal profile of network traffic and by defining a normal state of the system. Any deviation from the normal state indicates the presence of an anomalous activity in a network. Hence, semi-supervised and unsupervised machine-learning methods are frequently deployed in real-world security applications [13].

*Rule-Based Anomaly Detection*: In misuse detection, rules depict the strength of correlation between the conditions of the attributes and class labels. In the context of anomaly detection, the rules are the descriptors of normal profiles of users, application and system programs, and other resources in the computing and network infrastructures. An anomaly detection system is expected to raise an alarm of a potential attack if it observes any inconsistency among the current activities of the programs and the users with the established rules in the system. For an anomaly detection system to work effectively, it is critical to have an exhaustive set of rules working. The use of associative classification and association rules in anomaly-based intrusion detection systems is quite common. A number of propositions exist in the literature that has exploited the power of association rules in designing anomaly detection models [2, 14-15]. Anomaly detection systems using association rules broadly work in two steps. In the first step, effective data mining operations are carried out on the system and network audit data for identifying consistent and useful patterns of the behaviors of the programs and the users. In the second step, robust classifiers are inductively learned using the training dataset on the relevant features in the patterns to recognize any anomalous behavior in the system or in the network traffic. The concept of frequent episodes is presented in [16]. Lee and Stolfo utilized the concept of frequent episodes introduced in [16] to characterize the audit sequences occurring in normal data [2]. Based on the frequent episodes in the network, the authors designed a small set of rules that could effectively capture the frequent behaviors in those sequences. During the monitoring phase of the detection system, the event sequences that were found to violate the rules are identified as the anomalous events in the cyberinfrastructure.

*Fuzzy Rule-Based Anomaly Detection*: The anomaly detection systems working on the association rules use a deterministic value or an interval to quantify the rules. In such a scenario, the normal and anomalous records are separated by clearly defined and sharp boundaries in the *n*-dimensional feature space, where *n* is the number of features in the dataset. However, such a crisp separation poses a significant challenge in correctly detecting the normal audit records in situations where these normal data deviate from the established association rules by a small margin. This problem is handled by introducing fuzzy logic in designing the association rules, and thereby incorporating flexibility in the operations of rule-based anomaly detection systems.

Moreover, many of the features may be ordinal or categorical in nature, thereby making the design of association rules based on crisp and deterministic values of the features a well-neigh impossible proposition. Hence, the introduction of fuzziness in the association rules becomes mandatory. For example, a rule may contain the connection duration of a user's process by using the following expression, such as "connection duration = 3 min" or "1 min ≤ connection duration ≤ 4 min." Luo and Bridges investigated the fuzzy rule-based anomaly detection using real-world data and simulated data set [17]. The real-network traffic data were collected by the Department of Computer Science at Mississippi State University by *tcpdump* [18]. Four features were extracted from the data. These features were denoted as: SN, FN, RN, and PN. SN, FN, and RN denote, respectively, the number of SYN, FIN, and FST flags appearing in the TCP packet headers in the last two seconds. PN denotes the number of destination ports in the last two seconds. Three fuzzy sets were designed which were given names: LOW, MEDIUM, and HIGH. Each feature was divided into these three fuzzy sets. Fuzzy association rules were derived from the dataset based on the first three features of the data, and fuzzy frequency episode rules were designed for the last feature. Network traffic data in the afternoon of a given day was used in training of the model, and in deriving the fuzzy rules in the normal traffic data. The traffic data from the afternoon, evening, and night on the same day were used for testing and anomaly detection. For testing the model, a similarity function was used to compare the normal patterns with the anomalous patterns.

*Artificial Neural Networks*: Artificial neural networks (ANNs) allows for generalization in incomplete data and enables the detection of anomalous behavior in anomaly detection systems. The standard feed-forward multi-layer perceptron (MLP) with the ability of backpropagation of errors is particularly suited for carrying out anomaly detection. In the forward propagation phase, the ANN is trained on the training dataset. The data is fed into the network through the nodes in the input layer. The nodes at each layer are activated and their output passed on to the nodes in the next layer till the out values come of the nodes at the output layer. The output values produced by the output layer nodes are then compared with the desired or target values at the corresponding nodes. The difference between the actual output value and the target output value signifies the error at the node at the output layer. The error values are backpropagated through the links in the network from the nodes at the output layer back to the nodes at the input layer so that the weights in the links and the biases at the nodes can be updated. This process of forward and back propagation continues until the error values at the output nodes fall below a threshold value. At this point the training process completes. Ghosh et al. [19-20] and Liu et al. [21] applied ANNs in anomaly detection methods in computer networks.

*Support Vector Machines*: Support Vector Machines (SVMs) outperform ANNs in many situations as they have the ability to attain the global optimum state more efficiently and can control the model overfitting problem more effectively by fine-tuning the model parameters. SVMs can be gainfully deployed in anomaly detection by training them on datasets containing attack traffic and normal traffic. This is a supervised way of learning for SVMs. However, SVMs can also be applied effectively in an unsupervised way of identifying anomalous traffic in a network. Chen et al. used BSM audit data from the 1998 DARPA intrusion detection evaluation data sets and trained an SVM-based anomaly detection system using the dataset [22]. Hu et al. presented a comparative study on the performance of a robust support vector machine (RSVM) and a conventional SVM based on the nearest neighbor classification in separating normal traffic from attack traffic generated by various computer programs [23]. The results presented by the authors clearly showed that RSVMs had higher detection accuracy with a much lower value of false positives as compared with their conventional SVM counterparts. RSVMs also exhibited higher generalization ability in extracting information from noisy data.

*Nearest Neighbor-Based Learning:* Nearest neighbor-based machine-learning programs assume that the normal pattern of an activity displays a close displacement measured by a distance metric, while anomaly data points lay far from this neighborhood. K-nearest neighbor (KNN) method is a classification approach that uses a voting score among all the neighbors of a given data point in determining its class membership. The KNN learning-based anomaly detection method is effective only if the value of *k* is more than the frequency of occurrence any anomalous data in traffic audit dataset, and the Euclidean distance between the anomalous data groups from the normal data group is large in the *n*-dimensional feature space of the traffic dataset. In the literature, several anomaly detection approaches have been proposed using different variants of the basic nearest neighborhood-based classification method. These methods use different definitions of the nearest neighbour for the purpose of detection of anomalous traffic. Liao and Vemuri presented a KNN classifier model to classify the behavior of computer programs into two types – normal and anomalous [24]. In the proposed scheme, the behavior of a program was represented by the number of system calls made by the program. While every system call was treated as a word, the set of all system calls made by a program over its entire life span of execution was compiled as a document. The programs were subsequently classified into normal or anomalous classes using a KNN classifier constructed using document classification methods on the documents. The experiments were performed using the BSM audit data in the 1998 DARPA intrusion detection evaluation datasets. In the training phase, 3556 normal programs and 49 distinct system calls in 1 simulation day were used. The test audit data were scanned for programs to measure the distance. The distances were then sorted in the increasing order of their magnitudes and the top *k* scores were selected for the *k* nearest neighbors for each of the records in the test audit data. For the purpose of anomaly detection, a threshold value of the average of the top *k* distances for each record in the test dataset was determined. In their experiments, authors tried out different values of the threshold distance and the *k* values so as to determine the most optimal performance of the KNN classifier as depicted by its receiver operating characteristics (ROC) curve. The KNN algorithms were found to detect 100% of the attacks while keeping a *false positive rate* (FPR) at a very low value of 0.082% with *k* = 5 and a threshold value of 0.74.

*Hidden Markov Model and Kalman Filter*: Hidden Markov Model (HMM) considers transition properties of events. In network security applications it can be effectively deployed for detecting anomalous activities and events. In anomaly detection, HMMs can very accurately model the temporal variations in program behavior [25-27]. Before the deployment of an HMM in anomaly detection, the definition of a normal sate of activity *S* and a dataset of normal observable events *O* are to be decided upon. Starting from the initial state of *S*, and given a sequence of observations *Y*, the HMM searches for a sequence *X* that contains all normal states, and that has a predicted observation sequence which is most similar to *Y* with a computed probability value. If this computed probability value is smaller than a predefined threshold value, the sequence *Y* is assumed to have led the system to an anomalous state. Warrender et al. proposed an HMM-based anomaly detection model using publicly available datasets on systems calls from nine programs [25]. The datasets used were MIT LPR and UNM LPR [25]. An HMM with forty states was designed. These forty states represented forty systems calls that were present in all those nine programs. The HMM was designed in a fully connected manner so that transitions were possible from any given state to any other state in the model. The Baum-Welch algorithm was applied to fine-tune the parameters of the HMM using the training dataset [28]. The Baum-Welch algorithm works on the principles of dynamic programming and it is a variant of expectation maximization (EM) algorithm. The Viterbi algorithm was utilized to find out which choice of states maximizes the joint probability distribution given that trained parameter matrices of the HMM [29]. In other words, the Viterbi algorithm identifies the most likely state, given a dataset and a trained HMM model. The authors contend that for a well-designed HMM, a sequence of system calls that represents normal activities will lead to state transitions and output values which are highly likely; on the other hand, a sequence of system calls that represents an anomalous activity will lead to state transitions and output values that are unusual. Hence, in order to detect anomalous events in a network, it is sufficient to track unusual state transitions and abnormal output values. The experimental results indicated that the HMM could detect anomalous traffic efficiently and effectively with a low value of mismatch rate. In general, training of an HMM is a very time-consuming process as it requires multiple epochs (i.e., passes) through the records in a training dataset. Since all the transition probabilities corresponding to long sequences of state transitions are needed to be stored, training an HMM is a memory-intensive operation as well. Soule et al. presented an anomaly detection method in a large-scale data network [30]. The detection scheme analyzed the traffic patterns in a network, and computed the state the network using a Kalman filter. A Kalman filter is a set of mathematical equations that implements a predictor-corrector type estimation that is optimal [31]. The optimality here refers minimization of error covariance. The Kalman filter used in the anomaly detection filtered out the normal traffic state by comparing the predictions made by the current traffic state to an inference of the actual traffic state. The residual process is then analyzed for possible anomalies.

*Clustering-Based Anomaly Detection*: Supervised learning methods for the detection of anomalous activities in a network require prior labeling of the traffic types. However, it is very difficult to have prior labeling of audit data in real-world network environments. Signature-based detection suffers from this problem as carrying out a manual classification in a huge volume of network traffic to identify a small number of attack traffic records poses a significant challenge. Unsupervised learning-based anomaly detection methods do suffer from this drawback as these methods can work on unlabelled network traffic data. These methods attempt to detect malicious traffic in a network even without any prior knowledge about the traffic data labels. Unsupervised learning-based anomaly detection methods work under the following premise: in a network, characteristics of traffic are highly imbalanced - normal traffic constitutes a vast majority, while anomalous traffic represents a tiny minority. Moreover, attack traffic and the normal traffic exhibits similar statistical distributions in their respective group, while the distributions of the two groups are different from each other. Learning from an imbalanced data so that the anomalous and normal traffic can be categorized into two different clusters is the prime focus in unsupervised anomaly detection methods. Hence, cluster-based anomaly and outlier detection is the most fundamental approach in an unsupervised intrusion detection method. Portnoy et al. proposed a clustering-based anomaly detection method using DARPA knowledge discovery in databases (KDD) Cup 1999 dataset [32]. DARPA KDD Cup 1999 dataset consisted of a network traffic record of 4,900,000 data points. The dataset contained 25 different types of traffic - 24 attack types and 1 normal traffic. Each data point represented a set of extracted feature values from a connection record obtained between different IP addresses of hosts during a period of time in which attacks were simulated in a network. The authors observed that clustering with unlabelled data resulted in a lower detection rate of attacks than attack classification using a supervised learning method. However, unsupervised detection methods on unlabelled data can potentially detect unknown attacks through an automated or semi-automated process that cannot be done using supervised detection methods.

*Random Forests*: Random forests are powerful machine learning models based on ensemble approach. It builds multiple decision trees by randomly choosing a subset of features and then combines those decision tree results to arrive at a much more robust prediction. Due to their higher accuracy of prediction, random forests have been deployed in a variety of applications including multimedia information retrieval, network security and intrusion detection systems design. The algorithms used in random forests usually yield higher accuracy, and they work very efficiently on large datasets with high dimensional feature space. Traffic in a large network is an example of large volume high dimensional data, and such data can be very effectively classified in real-time by random forest-based classification approach. The use of random forest algorithms for detecting outliers in datasets containing network traffic without attack-free training data has been proposed in the literature [33-34].

*Other Machine Learning Methods in Anomaly Detection*: Other machine learning methods have been proposed for learning the probability distribution of data and in applying statistical tests to detect outliers. Eskin proposed a mixture probability model or normal and anomalous data based on expectation maximization (EM) algorithms [35]. Other statistical machine learning methods have been investigated in anomaly detection applications, such as mean and variance [4, 30], Hotelling's $T^2$ test and the Chi-square test [36, 37], Hellinger score [38], histogram density [39], Bayesian law [40], cumulative summation (CUMSUM) and statistical test [30]. Ye et al. used a series of probabilities techniques of anomaly detection, including decision tree, Hotelling's $T^2$ test, Chi-square multivariate test, and Markov chain in an information system for detecting intrusions [41]. Network-wide anomaly detection using principal component analysis (PCA) has been proved very effective [42-44]. Several studies have also found that a wide range of anomalies in networks can be detected by computing the entropy in the network flow and feature distributions [37,42,45]. Table 2 presents a summary of the anomaly-based detection schemes.

**Table 2: Anomaly -Based Detection Schemes**

| Detection Mechanism | Input Data Format | Detection Level | References |
|---|---|---|---|
| Statistical methods | Frequency of system calls, offline | Host | [36-37] |
| Statistical methods | TCP/IP packets, online | Network | [30], [38], [40] |
| Clustering algorithms | Frequency of system calls, online | Network | [25], [32-33] |
| Information theoretic | TCP/IP packets, offline | Network | [37],[42], [45] |
| Association rules | TCP/IP packets, offline | Host | [2], [14-16] |
| Fuzzy association rules | | | [17] |
| Kalman filter | TCP/IP packets, online | Network | [30] |
| Hidden Markov Model | Frequency of system calls, offline | Host | [25 -27] |
| Artificial Neural Network | Executables, offline | Host | [19-21] |
| Principal Component Analysis | Frequency of system calls, offline | Network | [42-44] |
| Support Vector Machine | TCP/IP packets, offline | Network | [22 -23] |
| K-Nearest Neighbors | Frequency of system calls, offline | Host | [24] |
| Random Forests | TCP/IP packets, offline | Network | [33-34] |

## 6. Machine Learning in Hybrid Detection

Since misuse detection systems work on matching already known attack signatures with the current events in a network, they usually have high detection rates and low false alarm rates. However, these systems cannot detect novel attacks. On the other hand, anomaly detection systems define normal sates in a network and then detect system states that significantly differ from the normal states. Any state that significantly differs from the normal state of the network indicates the possible event of an attack. The anomaly detection system can detect new attacks launched on a network. There is a challenge in the anomaly detection system design. If the normal state patterns do not significantly differ from patterns exhibited by any anomalous state, the attack state will go undetected. This leads to an increase in the false alarm rate. Hence, it is critical to design a normal state in such a way that while the detection rate is maximized, the number of false alarms does not exceed beyond an acceptable limit. If the normal state is too wide, then the detection rate will suffer. On the other hand, too narrow a normal state will lead to a high false alarm rate. The hybrid detection approach combines the adaptability and the powerful detection ability of an anomaly detection system with the higher accuracy and reliability of the misuse detection approach.

Designing an efficient and accurate hybrid detection system involves two critical issues: (i) the most ideal misuse or anomaly detection systems is to be first identified that can be integrated with anomaly detection systems, so that hybrid detection is possible, (ii) the two systems are to be integrated in the most optimal way so that the balance between the detection rate and false alarm rate is achieved while retaining the ability of detecting novel attacks.

The selection of misuse and anomaly detection systems for designing a hybrid detection system is dependent on the application in which the detection system is to be deployed. Following a combinational approach, the integration of an anomaly detection system with a misuse detection counterpart has been classified into four categories [46-47]. These types are: (i) anomaly-misuse sequence detection, (ii) misuse-anomaly sequence detection, (iii) parallel detection, and (iv) complex mixture detection. Figure 4. The complex mixture model is highly application-specific.

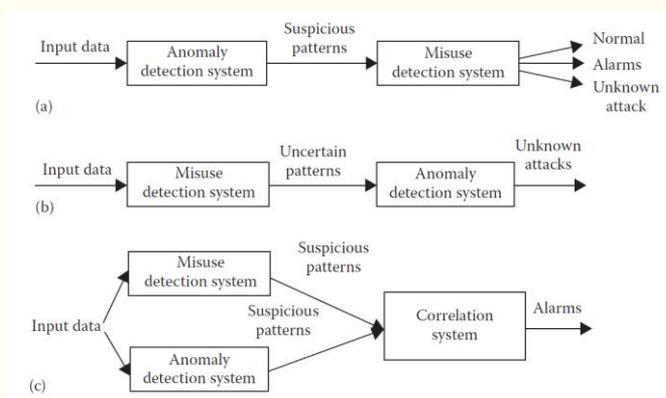

Figure 4: Three categories of hybrid detection systems. (a) Anomaly-misuse sequence, (b) Misuse-anomaly sequence, (c) Parallel detection system (Adapted from [43]

Barbara et al. presented a hybrid detection system on the principle of the anomaly-misuse sequence [48]. The proposed system which is known as audit data analysis and mining (ADAM), minimizes false alarm rates by not raising any alarm for those patterns that are not classified attacks by the misuse detection system. Misuse–anomaly sequence detection systems primarily focus on detecting novel attacks that are missed by the misuse detection module. The machine learning algorithms used by these hybrid detection models are mainly based on different variants of random forests [47, 49]. Anderson et al. proposed the design of a parallel intrusion detection system that provided a very accurate and robust detection decision by correlating the outputs of the misuse detection and the anomaly detection modules [50]. Agrawal et al. proposed an illustrative complex intrusion detection system [51]. The system worked on the AdaBoost algorithm of classification and both the misuse and the anomaly detection systems are trained on the training data simultaneously. The detection results on the test data are also presented separately for the misuse detection module and the anomaly detection module. Sen et al. proposed various architecture of complex detection systems based on cooperating agents [52-54]. The audit trails in the basic security module (BSM) of a Solaris system was exploited by Endler in designing a hybrid detection system [55]. An ANN-based hybrid detection system for detecting both signature-based and anomaly-based attacks is proposed by Ghosh & Schwartzbard [20]. Lee et al. presented a data mining-based hybrid intrusion detection system for identifying attack traffic from the audit data in a host [2]. Table 3 presents a summary of the hybrid detection systems discussed in this Section.

**Table 3: Hybrid Intrusion Schemes based on Machine Learning**

| Detection Mechanism | Input Data Format | Detection Level | References |
|---|---|---|---|
| Random forests | TCP/IP packets, online | Network | [46-47], [49] |
| Association rules | TCP/IP packets, online | Network | [48] |
| Association rules | Frequency of system calls, online | Host | [2] |
| Cooperating agents | TCP/IP packets, online | Network | [52-56] |
| Correlation | TCP/IP packets, online | Network | [50] |
| Clustering | TCP/IP packets, offline | Network | [51] |
| Statistical analysis and ANN | Sequences of system calls, offline | Host | [55] |
| ANN | TCP/IP packets, online | Network | [20] |

## 7. Conclusion

In this chapter, we have discussed various approaches to misuse and anomaly detection systems design using machine learning and data mining techniques. Some of the well-known systems in the literature have also been reviewed briefly. We have also discussed the pros and cons of various systems in context to their applications and deployment in real-world networks.

A fundamental challenge in designing an intrusion detection system is the limited availability of appropriate data for model building and testing. Generating data for intrusion detection is an extremely painstaking and complex task that mandates the generation of normal system data as well as anomalous and attack data. If a real-world network environment, generating normal traffic data is not a problem. However, the data may too privacy-sensitive to be made available for public research.

Classification-based methods require training data to be well-balanced with normal traffic data and attack traffic data. Although it is desirable to have a good mix of a large variety of attack traffic data (including some novel attacks), it may not be feasible in practice. Moreover, the labeling of data is mandatory with attack and normal traffic data clearly distinguished by their respective labels.

Unlike classification-based approaches which are mostly used in misuse detection, unsupervised anomaly detection-based approaches do not require any prior labeling of the training data. In most of the cases, the attack traffic constitutes the sparse class, and hence, the smaller clusters are most likely to correspond to the attack traffic data. Although unsupervised anomaly detection is a very interesting approach, the results produced by this method are unacceptably low in terms of their detection accuracies.

In a pure anomaly detection approach, the training data is assumed to be consisting of only normal traffic. By training the detection model only on the normal traffic data, the detection accuracy of the system can be significantly improved. Anomalous states are indicated by only a significant state change from the normal sate of the system.

In a real-world network that is connected to the Internet, an assumption of attack free traffic is utopian. A pure anomaly detection system can still be trained on a training data that includes attack traffic. In that case, those attack traffic data will be considered as normal traffic and the detection system will not raise an alert when such traffic is encountered in real-world operations. Hence, in order to increase the detection accuracy, attack traffic should be removed from the training data as much as possible. The removal of attack traffic from the training data can be done using updated misuse detection systems or by deploying multiple anomaly detection systems and combining their results by a voting mechanism.

For an intrusion detection system that is deployed in a real-world network, it is mandatory to have a real-time detection capability under a high-speed, high-volume data environment. However, most of the cluster techniques used in unsupervised detection require quadratic time. This renders their deployment infeasible in practical applications. Moreover, the cluster algorithms are not scalable, and they need the entire training data to reside in the memory during the training process. This requirement puts a restriction on the model size. The future direction of research may include studies on the scalability and performance of anomaly detection algorithms in conjunction with the detection rate and false positive rate. Most of the currently existing propositions on intrusion detection have not paid adequate attention to these critical issues.